\documentclass[fleqn]{article}
\usepackage{TAUP03_burgett_style}
\usepackage{amssymb}

\usepackage{graphicx}
\usepackage[figuresright]{rotating}

\newcommand{\AmS}{{\protect\the\textfont2
  A\kern-.1667em\lower.5ex\hbox{M}\kern-.125emS}}

\title{Searching for sources of the highest energy cosmic rays:
         low statistics, pitfalls, and possibile clues\thanks
{Talk presented at the Eighth International Workshop on Topics in Astroparticle and
                Underground Physics (TAUP 2003) September 5-9, 2003 at the University of
                Washington, Seattle, WA.}
}

\author{William S. Burgett\address{Institute for Astronomy, 
        University of Hawaii, 2680 Woodlawn Dr., Honolulu, HI 96821\\
        $^\mathrm{b}$Department of Physics, University of Texas at Dallas,
        P.O. Box 830688, Richardson, TX 75083-0688}%
        \thanks{email: burgett@ifa.hawaii.edu}
        and
        Mark. R. O'Malley$^\mathrm{b}$
        }
       
\begin{document}

\begin{abstract}
The clustering properties of the highest energy cosmic rays and their
correlations with candidate sources are re-examined using the most
recently available AGASA data and a rigorous correlation analysis. The
statistical methodology incorporates some important points not considered
in previous studies. Results include small angle clustering significances
consistent with, but somewhat less than, earlier findings, a possible
large scale anisotropy for events with energies $E \approx \; 5 - 8 \times
10^{19} \, \mathrm{eV}$, and no statistically significant cross correlations
with BL Lacertae or blazars. A marginally significant cross correlation
exists for events with energies $E > 8 \times 10^{19} \, \mathrm{eV}$
with a set of Abell clusters, but no definitive conclusion can yet be drawn
from this result.
\end{abstract}

\maketitle

\vspace*{-0.7cm}
\section{INTRODUCTION}

Investigating cosmic ray airshowers induced by extremely high energy cosmic
rays (EHECRs, $E \gtrsim 4 \times 10^{19} \, \mathrm{eV}$) remains an area of
intense interest because of an apparent paradox posed by the onset of the
Greisen-Zatsepin-Kuzmin (GZK) effect~\cite{greisen}. 
As is well known, the
paradox is that $\sim 15-20$ events have been detected
apparently having $E > 10^{20} \, \mathrm{eV}$ but with no currently obvious
astrophysical source within the local GZK sphere defined by the $50 - 100$ Mpc
distance over which cosmic ray protons lose a substantial fraction of their
initial energy through interactions with CMB photons~\cite{stecker}.
Due to this small number of events,
open questions include the existence of
statistically significant clustering/anisotropy, correlations
with known astrophysical source distributions (e.g., QSOs), 
the composition/charge of the primaries, and whether the production mechanism 
is top-down or bottom-up.

This presentation summarizes the analytical approach and
results contained in two papers~\cite{burgett1}.
Because of the three page limit imposed for the TAUP03 conference proceedings,
this paper is necessarily abbreviated,
informal, and contains only a few references.
However, further details concerning methodology, additional figures and tables, 
and more complete bibliographies can be found in our two papers mentioned above.

\section{ANALYTICAL APPROACH}

\begin{figure}
\centering
\includegraphics*[scale=0.5]{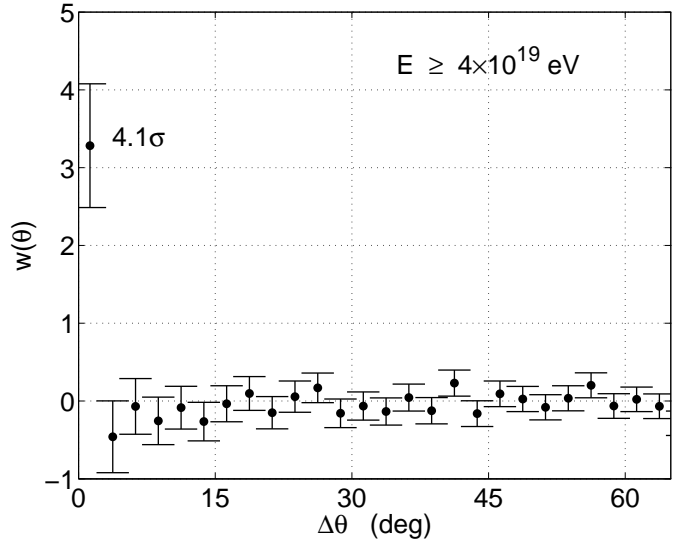}
\vspace*{-0.35in}
\caption{The two point angular correlation function for 60 AGASA events and 
$2.5^\circ$ bin widths.}
\label{corfcn_2pt5}
\end{figure}

For this study, we use the AGASA data for airshowers with 
estimated energies $E \geq 4 \times 10^{19} \, \mathrm{eV}$ and zenith angles $\leq 45^\circ$
contained in reference~\cite{takeda}. 
The results we obtain are based
on applying statistical estimators to Monte Carlo simulations of isotropically-distributed 
arrival directions convolved
with the observed AGASA detector acceptance.
The distribution of angular pair separations is analyzed with the Landy-Szalay (LS)
two point angular correlation function,
\vspace*{-0.3cm}
\begin{equation}
w(\theta) = DD/RR - 2DR/RR + 1 \; \; \mbox{,}
\end{equation}
\vspace*{-0.3cm}
where the ``D'' and ``R'' refer to a single angular position of an event in the data
or random catalogs, respectively, so that ``DD'' represents a pair separation between
two events in the data, etc.~\cite{landy}. Values of $w(\theta) > 0$ then indicate
pair correlations beyond those expected in uncorrelated distributions
(with a MC-derived significance). 
Figure~\ref{corfcn_2pt5} shows the results for bin widths of $2.5^{\circ}$ where the existence
of an excess number of pairs in the first bin over that expected from random projection is at
greater than the $4\sigma$ level.

\begin{figure}[t!]
\centering
\includegraphics*[scale=0.5]{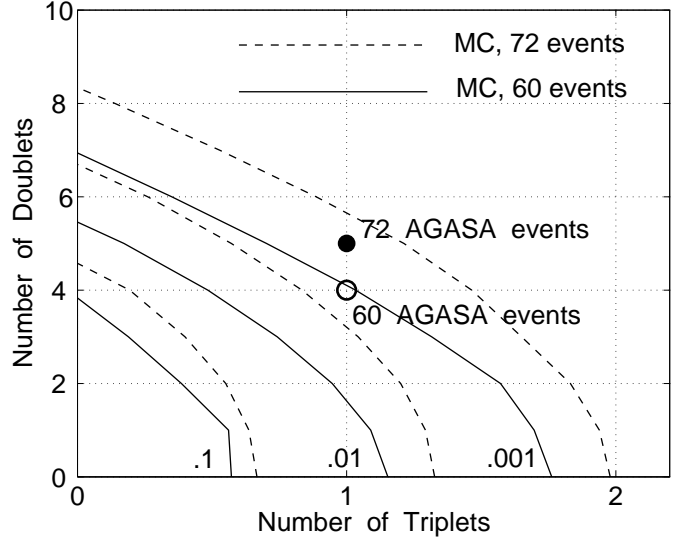}
\vspace*{-0.35in}
\caption{Joint inclusive probability contours for 60 and 72 AGASA event
sample sizes.}
\label{jp_60_72}
\end{figure}

Clustering probabilities are estimated directly from the simulations by utilizing a
counting scheme that identifies \textit{distinct} multiplet configurations (e.g., doublets,
triplets, etc.) instead of counting just pairs within a specified
opening (or correlation) angle. The reason for this is that higher order multiplets, such
as a triplet, generally have a lower probability for random occurrence than does the 
presence of the same number of distinct pairs. This can be seen from Figure~\ref{jp_60_72}
that shows the joint inclusive probability contours for various doublet-triplet configurations.

The associated LS cross correlation function is
\vspace*{-0.3cm}
\begin{equation}
\chi(\theta) = D_1D_2/R_1R_2 - (D_1R_2 + D_2R_1)/R_1R_2 + 1 \mbox{,}
\end{equation}
\vspace*{-0.3cm}
where the subscripts refer to the two samples being cross correlated, and values of
$\chi(\theta) > 0$ indicate the two samples are correlated (with a MC-derived significance). 
Equation (2) is used to investigate whether the AGASA sample can be separated into
statistically uncorrelated distributions partitioned by energy, and to search
for correlations between the AGASA sample and BL Lacertae, blazars, and Abell clusters
of galaxies.

Avoiding the pitfalls in cross correlating cosmic ray arrival directions
with catalogs of astrophysical objects requires accounting for\\
1. Selection effects such as absorption at low galactic latitudes,\\
2. Catalog completeness and source densities,\\
3. Intrinsic clustering/correlations between the candidate \textit{sources},\\
Many of the previously published studies finding correlations
with various types of objects have not fully considered these points. In our study, we have 
attempted to be as rigorous as possible.

\begin{figure}[t!]
\centering
\includegraphics*[scale=0.5]{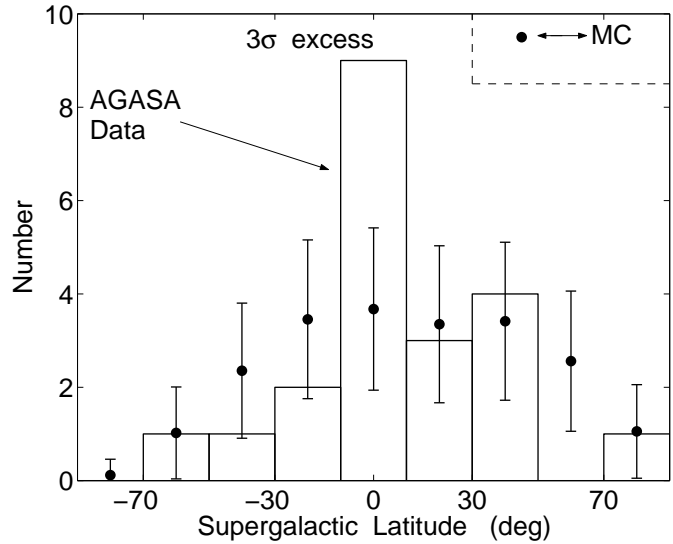}
\vspace*{-0.35in}
\caption{The supergalactic latitude distribution of the
$5 \leq E < 8 \times 10^{19} \, \mathrm{eV}$ partition.}
\label{sgb_part}
\end{figure}
\section{SUMMARY OF RESULTS}

\begin{figure}[t!]
\centering
\includegraphics*[scale=0.5]{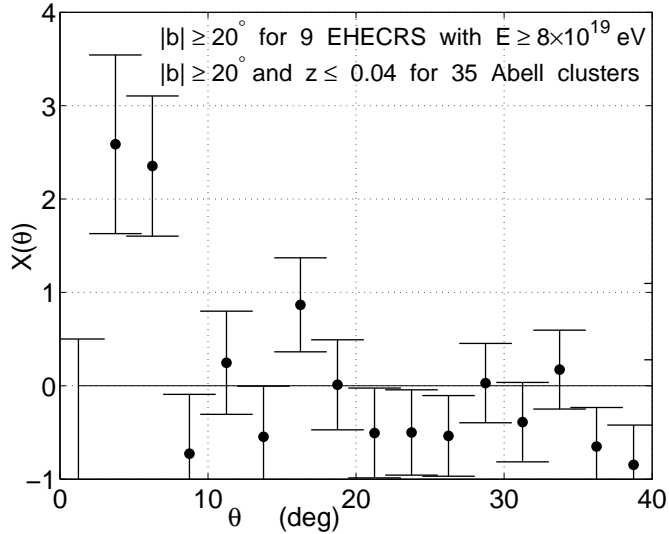}
\vspace*{-0.35in}
\caption{X-correlation between the highest energy AGASA EHECRS and
Abell clusters with $z \leq 0.04$; positive correlations exist to $z \leq 0.06$.}
\label{ag_abell_xcor}
\end{figure}

Our results can be summarized as follows:\\
1. Angular correlations exist on $\sim 2.5^\circ - 3^\circ$ scales at the
$\geq 4\sigma$ level (MC prob. $\sim 0.1\%$) with no departures from homogeneity
on scales $\gtrsim 4^\circ$,\\
2. Partitioning the AGASA sample by energy yields three apparently uncorrelated groups
with energies $E < 5\times10^{19} \, \mathrm{eV}$, $5 \leq E < 8 \times 10^{19} \, \mathrm{eV}$,
and $E \geq 8\times 10^{19} \, \mathrm{eV}$,\\
3. From Figure~\ref{sgb_part},
the partition with $5 \leq E < 8 \times 10^{19} \, \mathrm{eV}$ appears to be preferentially
aligned with the Supergalactic equatorial plane (i.e., in a plane containing a large fraction
of galaxies within our Local Supercluster) at the $0.1 - 0.6\%$ MC prob. level; 
the other two energy partitions are consistent with isotropic distributions,\\
4. None of the partitions exhibits a statistically significant correlation with BL Lacertae
or blazar sky (angular) positions drawn from three recent catalogs (Veron \& Veron-Cetty, 2002;
Padovani \textit{et al}., 1999 ; and Landt \textit{et al}., 2001),\\
5. The subset of events within the $E \geq 8 \times 10^{19} \, \mathrm{eV}$ partition having
galactic latitudes $|b| \geq 20^\circ$ appears to be correlated with $|b| \geq
20^{\circ}$, $80 \lesssim d \lesssim 250 \, h_{70}^{-1} \, \mathrm{Mpc}$ Abell clusters of galaxies
on angular scales of $2.5^\circ - 7.5^\circ$ at greater than the 99\% confidence level (the
maximum positive correlation appears between $3^\circ - 6^\circ)$~(see Figure~\ref{ag_abell_xcor}). 

\section{DISCUSSION}
We have no pretensions of having achieved definitive results. 
With the small data sample and low
statistics, the results could change considerably or disappear 
with only a few additional events.
However, the 
methodology is robust, and the internal consistency of the implied scenario is somewhat remarkable:
the mid-energy distribution ends at approximately the GZK ``cutoff'' energy, and is apparently
correlated with the luminous matter distribution in the Local Supercluster but not that
beyond. The high-energy partition may be isotropically distributed,
appears uncorrelated with the matter distribution of the Local Supercluster, but may be correlated 
with the matter distribution outside the Local Supercluster to distances as great as 
$250 \, h_{70}^{-1} \, \mathrm{Mpc}$.
Note that testing for energy partitions is independent of the subsequent cross 
correlation with other distributions.
Finally, penalty factors have been
considered and will be discussed in the forthcoming paper. 

\vspace*{0.5cm}
\noindent{\textbf{ACKNOWLEDEGEMENTS}}

The authors gratefully acknowledge the donations to the
        UT-Dallas Center for Theoretical and Interdisciplinary Physics
        by S. Bock, T. Ewing, H. Hancock, and M. Vick that made possible the
        presentation of this paper.


\newpage
\begin{thebibliography}{9}
\bibitem{greisen}
K.Greisen, PRL \textbf{16}, 748 (1966);
G.T. Zatsepin and V.A. Kuzmin, JETP Lett. \textbf{4}, 78 (1966).
\bibitem{stecker}
F.W. Stecker, PRL \textbf{21}, 1016 (1968);
C.T. Hill and D.N. Schramm, PRD \textbf{31}, 564 (1985);
V. Berezinsky and S.I. Grigor'eve, A\&A \textbf{199},
1 (1988);
S. Yoshida and M. Teshima, Prog. Theor. Phys. \textbf{89}, 933 (1993). 
\bibitem{burgett1}
W.S. Burgett and M.R. O'Malley, PRD \textbf{67}, 092002 (2003);
W.S. Burgett and M.R. O'Malley, ``Cross correlating AGASA EHECR
arrival directions with candidate sources and sites'', in preparation.
\bibitem{takeda}
M. Takeda \textit{et al}., ApJ \textbf{522}, 225 (1999);
N. Hayashida \textbf{et al}., e-print arXiv:astro-ph/0008102 (2000);
AGASA website.
\bibitem{landy}
S.Landy and A. Szalay, ApJ \textbf{412}, 64 (1993).
\end{thebibliography}
\end{document}